# A family of high-temperature ferromagnetic monolayers with locked spin-dichroism-mobility anisotropy: MnNX and CrCX (X=Cl, Br, I; C=S, Se, Te)

Cong Wang[†], Xieyu Zhou[†], Linwei Zhou, Ning-Hua Tong, Zhong-Yi Lu and Wei Ji[*]

*Beijing Key Laboratory of Optoelectronic Functional Materials & Micro-Nano Devices, Department of Physics, Renmin University of China, Beijing 100872, China*

[*] Corresponding author: Email: wji@ruc.edu.cn
[†] The authors contributed equally to this work.

**Abstract:**

Two-dimensional magnets have received increasing attention since $Cr_2Ge_2Te_6$ and $CrI_3$ were experimentally exfoliated and measured in 2017. Although layered ferromagnetic metals were demonstrated at room temperature, a layered ferromagnetic semiconductor with high Curie temperature ($T_c$) is yet to be unveiled. Here, we theoretically predicted a family of high $T_c$ ferromagnetic monolayers, namely MnNX and CrCX (X=Cl, Br and I; C=S, Se and Te). Their $T_c$ values were predicted from over 100 K to near 500 K with Monte Carlo simulations using an anisotropic Heisenberg model. Eight members among them show semiconducting bandgaps varying from roughly 0.23 to 1.85 eV. These semiconducting monolayers also show extremely large anisotropy, i.e. $\sim 10^1$ for effective masses and $\sim 10^2$ for carrier mobilities, along the two in-plane lattice directions of these layers. Additional orbital anisotropy leads to a spin-locked linear dichroism, in different from previously known circular and linear dichroisms in layered materials. Together with the mobility anisotropy, it offers a spin-, dichroism- and mobility-anisotropy locking. These results manifest the potential of this 2D family for both fundamental research and high performance spin-dependent electronic and optoelectronic devices.

bandgap semiconductor

# 1. Introduction

In the past decades, a plenty of methods, e.g. magnetic or non-magnetic dopants [1-7], vacancies [8, 9] and grain boundaries [10, 11], have been attempted to introduce long-range ferromagnetic orders in semiconductors. The long range order is, however, limited by the dopant-host hybridization [2, 12] that the highest $T_c$ was thus recorded at ~200 K [4, 13-15] in Mn-doped semiconductors and the mobility remains low. A $T_c$ of 340 K was observed in a new class of heavily Fe-doped materials [6, 16], which show metallic behaviors rather than semiconducting characteristics. It thus calls additional strategies for searching high $T_c$ ferromagnetic semiconductors. Recently, the re-discovery of mono- or few-layer two-dimensional (2D) materials, e.g. borophene [17], $MoS_2$ [18-20] and black phosphorus (BP) [21-23] boosted tremendous investigations on 2D semiconductors and their heterostructures [24-26]. Two intrinsic ferromagnetic semiconducting mono(bi)-layers, i.e. $CrI_3$ [27, 28] and $Cr_2Ge_2Te_6$ [29-31], were very recently demonstrated in experiments, although their Curie temperatures ($T_c$) are lower than 50 K. Thus, the high-$T_c$ ferromagnetic (FM) semiconductors are yet to be discovered although room-temperature FM metallic monolayers, e.g. $VSe_2$ [32, 33] and $Fe_3GeTe_2$ [34-36], were experimentally prepared or exfoliated, although with limited air-stability.

The lack of interlayer magnetic interactions at the monolayer limit enables a variety of interlayer antiferromagnetic (AFM) or weakly FM coupled materials to be under consideration for FM semiconductors, in which the constraint of strong FM interlayer coupling was eliminated. The $CrI_3$ mono- and bi-layers are exactly the case that they show an in-plane FM order and a weak interlayer AFM coupling below 50 K [28]. In a recent work, we found a strong electron doping (~1 $e$/Cr) to $CrS_2$ could lead this AFM metal to a FM semiconductor [37]. This doping could be realized by substituting Cr with Mn. Monolayer $MnS_2$ or $MnSe_2$ was predicted to be a FM semiconductor with a $T_c$ of ~200 K [38]. A MnSe$_2$ monolayer, with a measured room-temperature $T_c$, was recently fabricated by molecular beam epitaxy (MBE) although the monolayer strongly interacts with the substrate and is yet to be exfoliated [39]. Another route to realize the doping lies in replacing an S atom with a Cl atom forming a monolayer CrSCl. A Janus CrSCl monolayer is a FM semiconductor that is 85 meV/Cr more stable compared with other magnetic configurations [37], suggesting it to be a promising candidate for high-$T_c$ magnetic semiconductors.

Here, we theoretically predicted another structural form of CrSCl (Fig. 1a) and its analogues, e.g. CrCX and MnNX (C=S, Se, Te and X=Cl, Br, I). This form is energetically more stable than the Janus monolayers and does not have inherently existed unbalanced in-plane strain. This form, different from the hexagonal Janus [37], CrI$_3$ [28], RuCl$_3$ [40] Cr$_2$Ge$_2$Te$_6$ [29] and Fe$_3$GeTe$_2$ [34] monolayers, has two nearest and four second-nearest neighbors, twice the number of previous candidates in hexagonal lattices. Curie temperatures of these monolayers were predicted using Monte Carlo simulations with a third-nearest anisotropic Heisenberg (AH) model. The maximum $T_c$ values are over 240 K and near 500 K for semiconducting and metallic monolayers, respectively, which are much more superior to CrOCl [41], a lighter member of CrCXs with our predicted $T_c$ of 16 K. The bandgaps of these materials are primarily determined by the chalcogen atom, varying from nearly 2 to 0 eV, while eight of them are FM semiconductors and another four of them are FM metals with $T_c$ up to 492 K. In addition, this MnNX and CrCX family does not show strong structural anisotropy, but the electronic structures, carrier mobility and their resulting optical absorption are highly anisotropic and are locked together. All these results may boost experimental studies on this novel family of 2D magnetic monolayers with extraordinarily high $T_c$ values.

## 2. Methods

### 2.1. DFT calculations

Our density functional theory (DFT) calculations were performed using the generalized gradient approximation for the exchange-correlation potential, the projector augmented wave method [42, 43] and a plane-wave basis set as implemented in the Vienna ab-initio simulation package (VASP) [44, 45] and Quantum Espresso (QE) [46]. Dispersion correction was made at the van der Waals density functional (vdW-DF) level [47-49], with the optB86b functional for the exchange potential, which was proved to be accurate in describing the structural properties of layered materials [23, 50-54] and was adopted for structure related calculations. For energy comparisons among different magnetic configurations, we used either the PBE [55] or hybrid (HSE06) functional [56, 57], with the inclusion of spin-orbit coupling (SOC), based on the vdW-DF revealed structures. Density functional perturbation theory [58] was employed to calculate phonon dispersion (QE) and vibrational frequencies at the Gamma point (VASP). In VASP calculations, the kinetic energy cut-off for the plane-wave basis set was set to 700 eV for geometric and 400 eV for electronic structures calculations by the HSE06 functional. A $k$-mesh of 10×14×1 was adopted to sample the first Brillouin zone of the conventional unit cell of monolayer CrCXs and MnNXs. The phonon dispersion was obtained by Fourier interpolation of the dynamical matrices calculated using an 18×20×1 $k$-mesh and a 6×4×1 $q$-mesh with a plane-wave energy cutoff of 50 Ry. On-site Coulomb interactions to the Cr $d$ and Mn $d$ orbitals are considered with $U$ and $J$ values ranging from 3.89 – 4.40 eV and 0.80 – 1.25 eV, respectively, as revealed by a linear response method [59] and listed in the Supplementary Table S1. These values are comparable to the values adopted in modeling $CrI_3$ [60, 61] and $CrS_2$ [37]. The HSE06 functional already considers the exact exchange energy, the $U$ and $J$ correction does not apply to HSE06 calculations. The influences of different functionals and $U$, $J$ values were also discussed in the Supplementary Figs. S1, S2. More calculation details were provided in the Supplementary Materials.

**2.2. Curie temperature prediction**

Spin-exchange coupling (SEC) parameters were extracted based on a third-nearest Heisenberg model,

$$H = H_0 + J_1 \sum_{\langle ij \rangle} \mathbf{S}_i \cdot \mathbf{S}_j + J_2 \sum_{\langle\langle ij \rangle\rangle} \mathbf{S}_i \cdot \mathbf{S}_j + J_3 \sum_{\langle\langle\langle ij \rangle\rangle\rangle} \mathbf{S}_i \cdot \mathbf{S}_j.$$

Here, $J_1$, $J_2$ and $J_3$ represent the first-, second- and third- nearest couplings, respectively, as illustrated in Fig. 1a. We derived the spin exchange parameters by the total energy differences of the four magnetic configurations shown in the Supplementary Fig. S3. The magnetic energy contributions of these magnetic configurations in each magnetic unit cell write as

$$E_{FM} = \frac{N^2}{4} \times \frac{1}{2}(2J_1 + 4J_2 + 2J_3),$$

$$E_{AFM1} = \frac{N^2}{4} \times \frac{1}{2}(2J_1 - 4J_2 + 2J_3),$$

$$E_{AFM2} = \frac{N^2}{4} \times \frac{1}{2}(2J_1 - 2J_3),$$

$$E_{AFM3} = \frac{N^2}{4} \times \frac{1}{2}(-2J_1 + 2J_3),$$

where $N$ represents the unpaired spins on each Cr atom, which is chosen as 3 in our calculations.

Metropolis Monte Carlo simulations were carried out to predict Curie temperatures. A $J_1$-$J_2$-$J_3$ model was used to do the Metropolis Monte Carlo simulations, in which two nearest, and two next-next-nearest Cr (Mn) atoms were considered. This model is more advanced than the one only contains four next-nearest sites in a square lattice, as used in a previous work [41]. A 50×50 lattice was used for all simulations and each lattice point include 2 Cr (Mn) atoms. Simulation results revealed with the Ising model show high $T_c$ values from 271 to 1,179 K. However, the anisotropy was largely overestimated in the Ising model, which leads to the predicted $T_c$ values two or three times the exact values. We eventually used an anisotropic Heisenberg model in order to make the simulation results more sensible in comparison with experiments.

$$H = -(\frac{J}{2}\sum_{i,i'} \mathbf{S}_i \cdot \mathbf{S}_{i'} + \sum_i D_e(S_i^e)^2 + \frac{\lambda_e}{2}\sum_{i,i'} S_i^e \cdot S_{i'}^e).$$

The first term in the Hamiltonian describes the isotropic exchange while the final term is the anisotropic symmetric exchange between different sites. The term with D represents the easy axis single-ion anisotropy. Instead of the continuous flipping in the Heisenberg model, we fix the spin-flipping in only six directions: (1,0,0), (−1,0,0) ,

(0,1,0), (0,–1,0), (0,0,1), (0,0,–1); this simplification greatly speeds up simulation processes with a reasonable error of less than 10% as double checked in the Supplementary Fig. S4d. A hexagonal lattice was used for simulating the CrI3 monolayer, which yields a $T_c$ value of 65 K, very close to the analytical solution of 63 K; this verifies the reliably of the model. A $T_c$ value of 43 K was found through the anisotropic Heisenberg model, which is exceedingly closed to the experimental value of 45 K. In addition to MC simulations, we also employed renormalized spin wave theory (RSWT) to predict $T_c$ values. The details of the derivation and the numerical results of our RSWT calculations were provided in Supporting Information II.

**2.3. Carrier mobility estimation**

Phonon-limited carrier mobility in CrSeBr monolayers with a finite thickness $W_{\text{eff}}$ is expressed as [51, 62, 63]:

$$\mu_{\text{film}} = \frac{\pi e \hbar^4 C_{\text{film}}}{\sqrt{2}(k_B T)^{3/2} (m^*)^{5/2} (E_1^i)^2} F.$$

Here, $m^*$ represents the effective mass along the transport direction and $E_1$ is the deformation potential constant of the VBM (hole) or CBM (electron) along the transport direction, which is determined by $E_1^i = \Delta V_i/(\Delta l/l_0)$. Here $\Delta V_i$ is the energy change of the $i^{\text{th}}$ band under proper cell compression and dilatation (by a step 0.5%), $l_0$ is the corresponding lattice constant along the transport direction and $\Delta l$ is the deformation of lattice constant. Variable $C_{\text{film}}$ is the elastic modulus of the longitudinal strain in the propagation direction, which is derived by $(E - E_0)/V_0 = C(\Delta l/l_0)^2/2$; $E$ represents the total energy and $V_0$ represents the lattice volume at the equilibrium for 2D systems. A crossover function $F$ bridges the 2D and 3D cases, which is estimated by

$$F \equiv \frac{\sum_n \left\{ \frac{\sqrt{\pi}}{2}[1 - \text{erf}(\Omega(n))] + \Omega(n) e^{-\Omega^2(n)} \right\}}{\sum_n [1 + \Omega^2(n)] e^{-\Omega^2(n)}},$$

where

$$\Omega(n) \equiv \sqrt{\frac{n^2 \pi^2 \hbar^2}{2 m^* W_{\text{eff}}^2 k_B T}}.$$

The erf() represents an error function and the summation over integer is due to quantum confinement along the *z*-direction. Effective thickness of the film ($W_{\text{eff}}$) is expressed by:

$$\frac{1}{W_{\text{eff}}} = \int_{-\infty}^{+\infty} P_i(z)P_f(z)\mathrm{d}z = \sum_n \frac{\rho_i^n(z)}{N\Delta z} \cdot \frac{\rho_f^n(z)}{N\Delta z} \Delta z.$$

Here, $P(x)$ is the electron probability density along the *z* direction. We divided the space along the *x* direction into *n* parts by $\Delta z$. Variable $\rho^n(z)$ is the sum of the number of electrons $n^{\text{th}}$ region along the *z* direction. Here, *N* is the total number of valence electrons in the film, *i* and *f* represent equilibrium and deformed films, respectively. The electronic structures in carrier mobility elimination are all calculated with the HSE06 functional. We implemented these carrier mobility estimation methods in a computer package, "Renmin Mobility Calculator" (ReMoC). Please visit http://sim.phys.ruc.edu.cn/tools/ for details.

**2.4. Optical absorption spectra and conductivity calculation**

The absorption spectra were calculated from the dielectric function using expression [23, 51] $A(\omega)=\alpha(\omega)\cdot\Delta z$, where $\alpha(\omega) = \frac{\omega \mathrm{Im}\varepsilon}{cn}$ is the absorption coefficient, $n = \sqrt{\frac{\sqrt{(\mathrm{Re}\varepsilon)^2+(\mathrm{Im}\varepsilon)^2}+\mathrm{Re}\varepsilon}{2}}$ is the index of refraction, Re$\varepsilon$ and Im$\varepsilon$ are the real and imaginary parts of the dielectric function [64], respectively. $\omega$ is the light frequency, *c* is the speed of light in vacuum and $\Delta z$ represents the unit-cell size in the *z* direction. The conductivity tensor was calculated from the imaginary parts of the dielectric function as well, $\sigma_{ij} = \frac{\omega}{4\pi}\mathrm{Im}\varepsilon_{ij}$, where *i*, *j* represent the directions *x*, *y* and *z*. The electronic structures were obtained from the results unveiled using the PBE functional and the *k*-mesh was increased to $28 \times 32 \times 1$ in calculating dielectric functions. Enough conduction bands were considered and exciton effects were not considered in the optical properties calculations. Because the dielectric function is a tensor, the absorption spectra along the *x, y* and *z* directions were obtained separately. The energies of incident light of the horizontal axis in absorption spectra were shifted by the differences of bandgaps between the PBE+U-J (-SOC) and HSE06 (-SOC) results.

## 3. Results and discussion

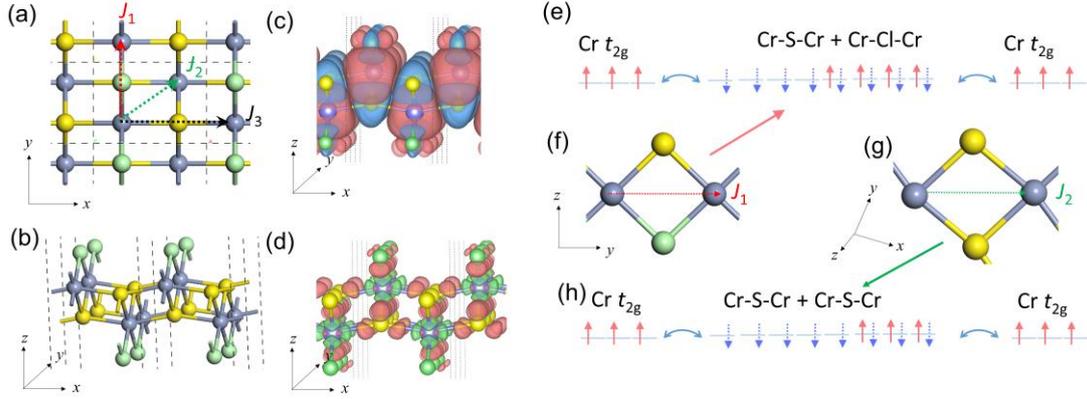

Fig. 1. Magnetic mechanism of CrCXs monolayers. (a), (b) Top and perspective views of a CrSCl monolayer in the orthogonal structure. Slate-blue, yellow and light green balls represent Cr, S and Cl atoms, respectively. The intralayer SEC parameters $J_1$, $J_2$, $J_3$ between Cr sites are represented with dashed arrows. (c) Perspective view of the spin charge density of CrSCl in the FM order plotted with an isosurface value of 0.001 $e$/Bohr$^3$. Red and blue isosurface contours correspond to charges polarized in up and down spins, respectively. (d) Perspective view of the atomic differential charge density of the CrSCl monolayer plotted with an isosurface value of 0.01 $e$/Bohr$^3$. Red and green isosurfaces correspond to the charge accumulation and reduction after Cr, S and Cl atoms bonding together, respectively. (e) Schematic of the FM double-exchange mechanism for spin-exchange through a Cr-S-Cr and a Cr-Cl-Cr channels (denoted $J_1$) as illustrated in (f), in which red up-oriented and blue down-oriented arrows represent electrons with different spins. By following the same scheme, panels (g) and (h) shows the atomic structure and spin-exchange coupling mechanism of the two Cr-S-Cr channels (denoted $J_2$), respectively.

Figs. 1a and 1b show a top- and a perspective-views of the fully relaxed atomic structure of CrSCl in the CrCXs form while its other less stable forms, e.g. 1T-Janus, are available in the Supplementary Fig. S5 and Table S2. Phonon dispersion spectra are also available in the Supplementary Fig. S6 indicating the stability of the CrCXs form. We used CrSCl as an example that it is comprised of perpendicularly oriented Cr-S/Cl rhomboid chains along $y$ and distorted Cr-S rectangular chains along $x$. The shortest Cr-Cr distance of 3.44 Å was found in the rhomboid chains, being bridged by an S and a Cl atoms. We denote the SEC parameter of this interaction as $J_1$ (red arrow). The second-nearest Cr-Cr interaction is bridged by two S atoms with a distance of 3.61 Å and a Cr-S-Cr angle of 91.5°, the SEC parameter of which is denoted as $J_2$ (green arrow). In addition, SEC parameter $J_3$ represents a nearly linear

Cr-S-Cr interaction (black arrow). Detailed structural information of CrCXs and MnNXs can be found in the Supplementary Table S3. The derivation details of these three parameters are available in the Section 2.

The FM state in all CrCXs and MnNXs is energetically more stable than other magnetic configurations regardless which functional is used and whether $U$ or/and $J$ is/are added, as shown in the Supplementary Fig. S3. By comparing their total energies, we derived $J_1$ = 0.90 meV, $J_2$ = 2.98 meV and $J_3$ = 1.26 meV for CrSCl. Exact SEC parameters and predicted Curie temperatures are listed in Table 1. Here, we used a classical magnetic moment $S$=3/2 according to the DFT value of roughly 3 $\mu_B$ per Cr atom. Although $J_2$ represents a 0.17 Å longer distance than $J_1$, the coupling strength of $J_2$ is triple that of $J_1$. Replacement of Cl with Br or I substantially enlarges $J_1$ and slightly enlarges $J_2$ and $J_3$ in the S/Se-series, leading to nearly comparable $J_1$ and $J_2$ for the I-series monolayers. The Te-series is fairly different from the S/Se-series that the comparable $J_1$ and $J_2$ values increase from Cl to I and $J_3$ drops from nearly 16 meV to a small negative value (–1.9 meV), leading to small $J_1$ and $J_2$ and pronounced $J_3$ in CrTeCl and three times larger $J_1$ and $J_2$ and negative $J_3$ in CrTeI. All these three parameter gradually enlarge from Cl to I in the MnN series and those of MnNI is significantly large.

Either spin density (Fig. 1c) or atomic differential charge density (aDCD, Fig.1d) shows the origin of magnetism from the Cr $t_{2g}$ orbitals where three Cr 3$d$ electrons fill in. The aDCD indicates charge reduction on Cr $e_g$ orbitals forming sigma bonds with chalcogen or halogen 3$p$ orbitals. The spin density indicates the spin-up (red) was primarily originated from Cr $t_{2g}$ orbitals and partially from halogen $p_z$ orbitals. The spin-down (blue) component are mainly contributed by all three chalcogen $p$ and partially by both halogen $p_x$ and $p_y$ orbitals, suggesting the super-exchange FM coupling along the $J_1$ or $J_2$ direction bridged by S or Cl atoms.

Figs. 1e–1h summarize a diagram of the orbital filling and their magnetic interactions, which were compellingly supported by the orbital-decomposed bandstructures (the Supplementary Fig. S7). In particular, two Cr atoms are bridged by a Cl and an S atom for the $J_1$ interaction (along the $y$ direction), which involves

two of six $sp^3d^2$ hybridized orbitals for each Cr and two orthogonal $p_{x/y}$ orbitals for each Cl or S, totally eight orbitals. In a valence band picture, each Cr orbital is originally filled by 3/2 electrons while 5/3 and 4/3 electrons for each Cl and S orbitals, respectively. These eight orbitals are thus filled by 6*e* from Cr, 10/3 *e* from Cl and 8/3 *e* from S, totally 12*e*. Therefore, four states are only filled by spin-down electrons (Figs. 1e and 1f), which is consistent with the spin density plotted in Fig. 1d. These orbitals offer four channels of virtual hopping for the spin up electrons of the Cr $t_{2g}$ orbitals; this strongly favors FM coupling and is consistent with the Hund's rule. This picture also explains the always slightly stronger $J_2$ than $J_1$ that the Cr-S-Cr coupling (Figs. 1g and 1h), having a 2/3 *e* reduced number of filling electrons, offers an additional partial channel for FM coupling to the Cr-Cl-Cr coupling. It also suggests that a reduced number of the filling spin-down electrons of the bridging atoms may open more hopping channels and thus enlarge the value of $T_c$, which was elucidated by the 492 K $T_c$ of the MnNI monolayer (see Table 1).

In a local moment picture, the linear Cr-S-Cr coupling ($J_3$) usually favors AFM. Here, the hybridization of chalcogen $p_z$ with Cr $d_{z^2}$ orbitals form delocalized bonding and antibonding states (see Fig. 2 and Supplementary Fig. S8). The bonding state is occupied and itinerant electrons of this state dominate and mediate a FM coupling between two adjacent Cr cations, which is reinforced by kinetic energy gains [65]. Therefore, the linear $J_3$ gives rise to FM coupling in the CrS/Se and MnN series. In the Te series, the Cr $d_{z^2}$ orbitals become partially occupied and dominate the bandstructures around the Fermi level thus leading to a strong itinerant FM in the monolayer limit; this explains the fairly large FM $J_3$ in CrTeCl. The occupation of the antibonding conduction bands gradually reduces from CrTeCl to CrTeI (the Supplementary Figs. S9g–S9i). Therefore, the super-exchange AFM coupling is eventually overcome the FM coupling, leading to a small AFM $J_3$ for CrTeI.

Monte Carlo simulations were performed with an anisotropic Heisenberg (AH) model and a three-nearest Ising model (see the Supplementary Fig. S4 and Methods for details). Here, the AH model considers both on-site and spin-spin anisotropies, which play a key role when the thickness of a layered materials reduces to its 2D limit.

The on-site anisotropy is primarily a result of spin-orbit coupling (SOC). For spin-spin interactions, we used a simplified model where the interactions along two hard magnetization axes were averaged. Table 1 shows the single-ion magnetic anisotropy energies (MAEs) of all considered CrCXs/MnNXs. The easy axes of CrSCl, CrTeBr, CrTeI, MnNCl and MnNBr were found parallel to the *z* direction. This group contains direct and indirect bandgap semiconductor and metals with both highly dispersive and nearly flat bands around the Fermi Level. Table 1 also shows the RSWT predicted $T_c$ values. The RSWT simulations give nearly identical $T_c$ values to the MC results for CrSBr and CrSeI. For other layers, The RSWT results show substantially higher $T_c$ values compared with MC calculations, except for CrTeI where RSWT predicts a value of 86 K while MC gives a value of 139 K. We found the use of HSE leads to larger FM spin-exchange parameters and reinforced spin-spin and on-site FM couplings for magnetic anisotropy (the Supplementary Table S4); this thus results in higher transition temperatures. In light of this, our predicted transition temperatures using MC are rather conservative and the measured $T_c$ values might be even larger than our predicted values.

Table 1 Magnetic properties of the 12 monolayers, including the intralayer spin-exchange coupling parameters $J_1$, $J_2$, $J_3$, anisotropic spin-spin exchange parameter $\lambda$ and easy axis single ion anisotropy $D$, easy axis direction, Curie temperatures predicted using Ising model, anisotropic Heisenberg (AH) model and RSWT, bandgaps without on site Coulomb $U$ and exchange $J$ (w/o $UJ$), bandgaps with and without spin-orbit coupling revealed with the HSE06 functional.

|  | Exchange parameters (meV/Cr) | | | | | Easy axis | $T_c$ (K) | | | Bandgap (eV) | |
| --- | --- | --- | --- | --- | --- | --- | --- | --- | --- | --- | --- |
|  | $J_1$ | $J_2$ | $J_3$ | $\lambda$ | $D$ |  | Ising | AH | RSWT | HSE/ -SOC | PBE+$U$-$J$/ -SOC |
| CrSCl | 0.90 | 2.98 | 1.26 | 0.01 | –0.02 | $z$ | 271 | 108 | 320 | 1.86 /1.85 | 1.40 / 1.52 |
| CrSBr | 1.66 | 3.09 | 1.52 | –2×10$^{-3}$ | 0.02 | $x$ | 313 | 127 | 124 | 1.76 /1.68 | 1.36 / 1.45 |
| CrSI | 2.49 | 3.07 | 1.69 | –4×10$^{-4}$ | 0.06 | $y$ | 352 | 146 | 181 | 1.21 /1.09 | 0.89 / 1.01 |
| CrSeCl | 1.32 | 3.40 | 0.56 | 0.02 | –0.01 | $y$ | 284 | 118 | 213 | 0.85 /0.81 | 0.57 / 0.86 |
| CrSeBr | 2.09 | 3.52 | 0.76 | 0.01 | –4×10$^{-3}$ | $y$ | 326 | 135 | 227 | 0.96 /0.89 | 0.64 / 0.87 |
| CrSeI | 3.12 | 3.67 | 1.18 | –1×10$^{-3}$ | 0.02 | $y$ | 391 | 164 | 164 | 0.74 /0.55 | 0.46 / 0.62 |
| CrTeCl | 1.45 | 1.42 | 16.28 | 0.05 | 0.15 | $x$ | 589 | 248 | 377 | -* / -* | -* / -* |
| CrTeBr | 2.91 | 2.68 | 4.61 | 0.48 | –0.75 | $z$ | 448 | 187 | 582 | -* / -* | -* / -* |

| | | | | | | | | | | |
|---|---|---|---|---|---|---|---|---|---|---|
| CrTeI | 4.56 | 4.13 | −1.90 | −0.10 | 0.55 | z | 304 | 139 | 86 | -* / -* | -* / -* |
| MnNCl | 5.53 | 4.15 | 3.35 | 0.02 | 0.07 | z | 600 | 238 | 376 | 0.44 /0.68 | 0.47 /0.46 |
| MnNBr | 5.66 | 3.68 | 5.41 | 0.03 | 0.02 | z | 652 | 261 | 469 | -* / 0.23 | 0.32 / 0.33 |
| MnNI | 14.14 | 5.71 | 8.19 | −0.03 | 0.93 | x | 1179 | 492 | 741 | -* / -* | -* / -* |
| $CrI_3$ | 1.62 | N/A | N/A | 0.08 | 0.06 | z | 65 | 43 | 52 | | N/A |
| CrOCl | −0.02 | 0.07 | 1.46 | −2×10$^{-3}$ | 0.02 | z | 29 | 16 | 12 | 3.20 / | 2.47 /2.54 |

* stands for metal

In terms of metallic layers, MnNI has the highest $T_c$ of 492 K and the lowest one of 139 K was found in CrTeBr. The bandgap of MnNBr is less conclusive that the HSE and PBE-UJ calculations suggest different results. These metallic FM monolayers, serving as FM metals in transitional magnetic devices, offer flexibility and largely reduced thickness in 2D magnetic devices. The $T_c$ values of semiconducting monolayers were lower than those of metallic layers. In particular, the highest $T_c$ of 238 K (600 K with the Ising Model and 376 K with RSWT) was found in MnNCl (Table 1) with the easy axis along z while the lowest $T_c$ is 108 K for CrSCl (AH model value) or 124 K for CrSBr (RSWT value). These values are roughly five times and twice the measured value of 45 K [28] and this-work predicted value of 43 K for $CrI_3$. Here, a $T_c$ over 200 K approaches the room temperature and is subject to further substrate induced enhancement, as found in the $MnSe_2$ [38, 39] and $Cr_2Ge_2Te_6$ [29] cases, and the doping enhancement as realized in $Fe_3GeTe_2$ [34-36].

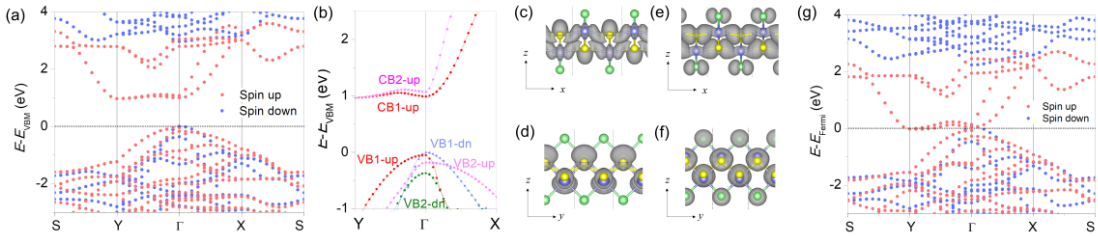

Fig. 2 Electronic structures of CrSeBr. Spin-resolved, i.e. red for spin-up and blue for spin-down, electronic band structure of CrSeBr were calculated with the HSE06 functional (a). A simplified bandstructure of CrSeBr around the G point was plotted in (b). Wavefunction norms of spin-up CB2 along G-X and G-Y were shown in (c) and (d), respectively, while those for spin-up VB1 were depicted in (e) and (f). The isosurface value was set to 0.001 $e$/Bohr$^3$. Spin-resolved electronic band structure of CrTeI calculated with the HSE06 functional was shown in (g).

Table 1 suggests that the S and Se series, CrTeI (inconclusive) and MnNCl are semiconductors with varying bandgaps from 1.85 to 0.55 eV (HSE-SOC) / 1.52 to 0.33 (PBE-UJ-SOC), among which CrSeCl is of a near direct-bandgap (direct- and indirect-bandgaps differ by 0.5 meV) and CrSCl also offers a small difference of 6 meV. According to a private communication with Yu Ye from Peking University, PBE-UJ-SOC result, with our linear-response derived $U$ and $J$ values, for the bandgap of CrOCl layers appears to be closer to the experimental value than that of HSE. In light of this, our $U$ and $J$ values should be very close to the set fitted using bandgaps. However, we cannot rule out the role of defects in the bandgap measurements, we thus used the HSE values in our following discussion on electronic and optical properties, which are independent from $U$ and/or $J$ values and were found comparable with experimental values in 2D layers. Fig. 2a depicts the bandstructure of CrSeBr because of its moderate bandgap (0.89 eV, in the infrared range), high predicted $T_c$ and a representative bandstructure of CrCXs. Bandstructures of other monolayers calculated with different functionals were shown in the Supplementary Figs. S9–S12.

Strong anisotropy, nearly linear dispersion and a roughly flat band are more clearly shown in a simplified bandstructure of CrSeBr (Fig. 2b), which contains two spin-up CBs, two spin-up and two spin-down VBs, respectively. The spin-up CB2 (pink) is comprised of Cr $d_{z^2}$ and Se/Br $p_z$ orbitals (Supplementary Fig. S7b-d), which exceptionally form extended Cr-Se-Cr channel states along the $x$ direction (Fig. 3c) but highly localized along the $y$ direction (Fig. 3d), giving rise to a quasi-1D electronic state in a 2D orthogonal lattice. Note that we define the $d_{z^2}$ or $p_z$ direction along the $x$ direction and $p_x$, $p_y$ and $d_{x^2-y^2}$ along the Cr-X (X=Cl, Br and I) bonding directions (see Supplementary Fig. S7a). This quasi-1D state offers a small effective mass of 0.06 $m_0$ along G-X but a rather large effective mass of 1.18 $m_0$ along G-Y, leading to the mobility (6.10 × 10$^3$ cm$^2$/Vs) for spin-up CB2 along $x$ 68 times that along $y$ (0.09×10$^3$ cm$^2$/Vs) (Supplementary Table S5). State spin-up CB1 (red) shares the same feature but its wavefunction is more localized along either $x$ or $y$ (Supplementary Figs. S8f and S8h), consistent with the larger effective masses of 0.40 $m_0$ and 7.1 $m_0$, respectively. The position of the spin-down component of CB sites over 1 eV higher than the spin-up CBM, while it also slightly depends on $U$ values in PBE-UJ calculations (see the Supplementary Fig. S1).

Spin-up VBs are rather interesting that the anisotropy of effective masses was found in either spin-up-VB1 (red) and -VB2 (pink), both of which are composed of Se/Br $p_z$ and Cr $d_{z^2}$ orbitals (Supplementary Fig. S7e-g). The spin-up-VB1 appears a nearly mirror analogue to that of spin-up-CB1 with respect to the gap around the G point, namely $m_x$=0.06 $m_0$ and $m_y$=1.30 $m_0$ (Supplementary Table S5), which shares the same mechanism of the CB case (Figs. 2e and 2f). The mass anisotropy of spin-up-VB2, i.e. 0.53 $m_0$ ($y$) and 2.66 $m_0$ ($x$), is reversal to spin-up-VB1 along the $x$ and $y$ directions. A similar behavior was found for those two spin-down VBs (blue and green) with less pronounced anisotropy (Supplementary Fig. S8 and Table S5). In CrSeBr, spin-down-VB1 serves as the highest VB, which may change in other CrCXs, e.g. in CrSCl, where the monolayer becomes a nearly direct-bandgap half-semiconductor (Fig. 3 and Supplementary Fig. S9).

The giant effective masses found along either direction for these bands imply likely strong correlation of VBs or doped CBs, which is, most likely, more pronounced in CrTeI (Fig. 2g). It shows a nearly flat-band along G-Y near $E_F$, which is even flatter and closer to $E_F$ with the inclusion of SOC (Supplementary Fig. S10i). Such a flat band mixed with two spin components might suggest emerging physical phenomena with strong correlation. In terms of MnNXs, they contain band crossings along G-Y with likely band inversions. The semiconducting MnNCl is a result of interaction induced gap-opening (Supplementary Fig. S10e). Other semiconducting layers, except MnNCl, show similar anisotropic features and share the same mechanism of anisotropy (Supplementary Fig. S9). This general spin-dependent mobility anisotropy suggests spin-up electron carriers moving much faster along the $x$ direction than along the $y$ direction and the reversal for hole carriers. If an off-axis in-plane electric field is established, spin accumulation might be observed in the more localized $y$ direction, which may potentially be used in transferring spin torques. All band structures and detailed energy levels of VBM and CBM of CrCXs and MnNXs are available in Fig. 3 and Supplementary Fig. S9.

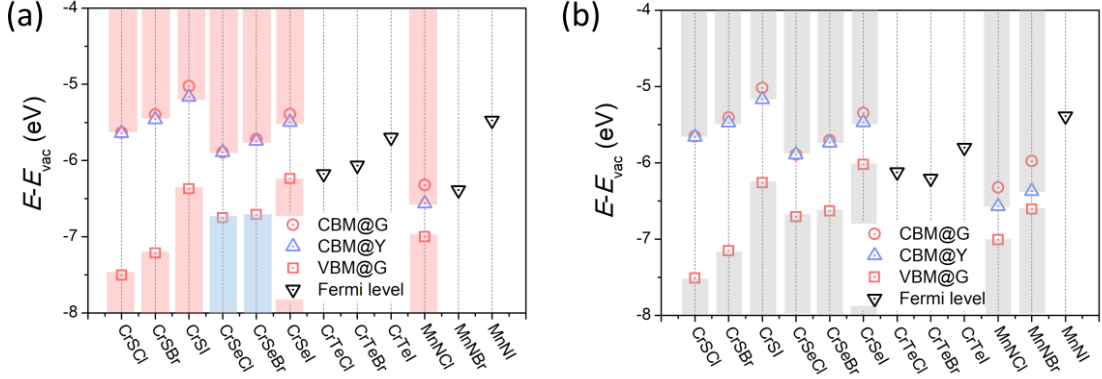

Fig. 3. Energy positions of CBM, VBM or the Fermi level of CrCX monolayers. The energy positions were obtained from the HSE06 (a) and HSE06-SOC (b) results, respectively. The vacuum level was set as the reference zero. Red and blue bars indicate the spin-up and -down components, respectively. In the MnNCl/MnNBr monolayer, the blue triangle shows the position of its CBM which sits at a point between G and Y. The Fermi levels in all metallic monolayers were labeled with black triangles.

Figs. 3a and 3b plots the positions of CBMs and VBMs calculated without and with SOC, respectively. The energy levels of CBM at G and Y have tiny differences below 0.15 eV and an 1% in-plane strain can induce a transition between indirect and direct bandgap. It is exceptional that the energy positions of the CBMs for all semiconducting CrCXs and MnNCl monolayers are rather deep, from –5.17 to –6.57 eV, suggesting the anisotropy of CBs could be feasibly utilized in practical devices. The band alignments indicate the possibility of constructing type-I (e.g. CrSeI/CrSI), type-II (e.g. CrSeCl/CrSeI) and type-III (e.g. CrSeI/MnNCl) heterostructures. The type-III heterostructures are of particular interest that they may be employed to build Dirac-source devices with the subthreshold swing smaller than 60 meV/decade [66].

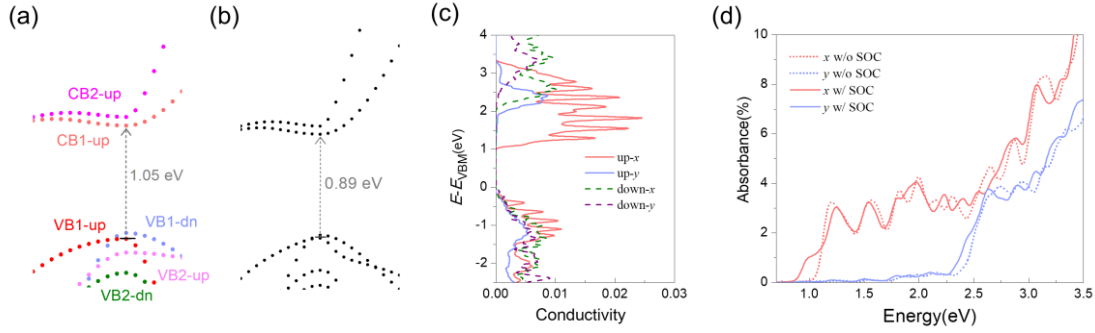

Fig. 4. Optical conductivity and absorbance. (a), (b) Sketch map of CB1, CB2, VB1 and VB2 of CrSeBr around the G point calculated with the HSE06 (a) and HSE06-SOC (b). (c) Spin resolved optical conductivities of different incident light polarization directions. (d) Absorbance of incident light polarized along the *x* (red) and the *y* (blue) directions in an energy range from 0.7 to 3.5 eV.

Fig. 4a shows a sketch-map of the bands around the G point of the CrSeBr monolayer. The CB energies at G and Y are nearly degenerated. Since the halogen atoms strongly affect the interactions along the *y* direction, heavier halogen atoms lead to more extended wavefunctions and thus show stronger band dispersions; this gives rise to flatter CBs and large splitting of spin-up-CB1 and -CB2 in CrSCl than those in CrSI (Supplementary Figs. S9 and S11). An exchange-induced enhancement of CB splitting at G was found when comparing results using the hybrid HSE06 functional with those of PBE (Supplementary Figs. S9 and S11). In addition, the inclusion of spin-spin exchange *J* or on-site U term in HSE calculations, as a result of overestimated FM exchange, enlarges the splitting of the two CB at the G point (Supplementary Fig. S2), resulting in a lower energy of CB at G than Y, which, we believe, is an artefact in the groundstate but might be valid under certain external fields. These results highlight the importance of exchange interactions in obtaining direct bandgap at the G point in CrCXs.

The optical conductivity [67] (Fig. 4c) and the light absorbance [51, 64] (Fig. 4d), explicitly show a spin-selective linear dichroism. In particular, only the up-spin component has absorbance in an energy range from 1.05 eV (transition from spin-up-VB1 to -CB1) to 2.27 eV while the down-spin component becomes excitable by light with even higher energies; this allows to generate pure spin-polarized photo-current using light linearly polarized in certain directions. In this range, light linearly polarized along the *x* direction has much higher absorbance (4.26 % per layer at 2.00 eV) than that along the *y* direction (0.31 % per layer at 2.00 eV), showing a spin-selective linear dichroism, in other words, a spin-dichroism locking. This spin-locked linear dichroism is thus substantially different common spin-dependent optical excitations by left- or right-hand circularly polarized light in magnetic semiconductors and the circular dichroism led by valley physics in transitional metal dichalcogenides (TMDs) [68-74], as well as the linear dichroism induced by wavefunction asymmetry [20].

Although the absorption of CrSeBr initializes in the infrared range, this energy range is tunable to lower or higher energy range by substituting S or Cl with Se, Br or I atoms since it is highly relevant with the bandgaps. In the presence of Se, Br and I atoms, strong SOC eliminates symmetry forbidden rules of different spin components, which shifts the absorption edge of CrSeBr from 1.05 to 0.89 eV (Fig. 4d) since the both spin channels could be excited to the CB states. Therefore, we infer CrSeI (with a gap of 0.55 eV, equivalently 2,255 nm) may have better absorbance among these CrCXs. In addition, this dichroism suggests CrCX monolayers may be used to construct spin-optical selectors. In particular, pure spin-up photo-electrons, moving much faster in the *x* direction, were excitable only by linearly polarized lights along the same direction while the excitations of spin down electrons are forbidden; this might shed light on generating pure spin-polarized photo-current by only linear polarized incident light.

In summary, we discovered a family of ferromagnetic semiconducting and metallic monolayers, namely CrCXs and MnNXs, which share the same structure. Their $T_c$ values are predicted up to 492 K with strong intra-layer and weak inter-layer magnetic couplings. Both localized and itinerant electrons contribute to the formation of the FM ordering. The strengths of these two competing coupling mechanisms are tunable by element substitutions. A reduced number of filling electrons to the chalcogen and halogen atoms was found a key factor for stronger FM couplings, which enlightens a strategy for searching other high $T_c$ FM monolayers. Given over 100 K $T_c$ for CrSCl, it turns out the strong SOC is not paramount in obtaining high $T_c$ in 2D layers [75] while spin-spin interaction or orbital anisotropy does offer magnetic anisotropy, which opens an avenue for searching magnetic monolayers. Additional high $T_c$ indicators include greater local magnetic moment (usually three $t_{2g}$ electrons) and larger neighboring numbers.

Strong anisotropy is a key characteristic of this monolayer family. The spin-up VBs and CBs of semiconducting CrCXs show highly anisotropic electron and hole effective masses and mobilities that the mass ratio is up to 22 and the mobility ratios are near 670 between the $x$ and $y$ directions. Strongly dispersive S/Se $p_z$ and Cr $d_{z^2}$ states along $x$ and nearly flat-band Cr $d_{z^2}$ states along $y$ were exceptionally found in CrCXs, resulting in the coexistence of nearly free-electron and highly correlated states. Here, the chalcogen atoms do not include O since the O-Cr polarization is much stronger than those in other bonds leading to more localized states. In addition to the previously revealed purely circular and linear dichroisms found in $MoS_2$ [68, 69], BP [23, 76, 77] and other 2D materials [70-74], we found a spin-selective (locked) linear dichroism in 2D CrCX layers, showing a spin-dichroism-mobility locking effect. Only the spin-up component could be excited with linearly polarized light along the $x$ direction and the excited spin-up electrons moving much faster along the same direction. Note that the energy difference between interlayer FM and AFM configurations for each CrCX or MnNX is roughly 1 meV/Cr or 1 meV/Mn (not shown here) and the bulk form of CrSBr was experimentally synthesized [78]. These facts suggest the feasibility of experimental investigations and manipulations of the CrCXs and MnNXs monolayers. All these results compellingly indicate that the CrCXs / MnNXs family is a novel interesting category of FM monolayers for either spintronics or optoelectronics, which is experimentally accessible and is of high potential in applications.

**Conflict of interest**

The authors declare that they have no conflict of interest.

**Acknowledgments**

This work was supported by the National Natural Science Foundation of China (11274380, 91433103, 11622437 and 61674171), the Fundamental Research Funds for the Central Universities of China and the Research Funds of Renmin University of China (16XNLQ01), the Strategic Priority Research Program of Chinese Academy of Sciences (XDB30000000). C.W was supported by the Outstanding Innovative Talents Cultivation Funded Programs 2017 of Renmin University of China. Calculations were performed at the Physics Laboratory of High-Performance Computing of Renmin University of China and Shanghai Supercomputer Center.

**Author contributions**

Wei Ji designed the research. Cong Wang performed the calculations of the geometry structures, electronic structures and magnetic coupling parameters. Xieyu Zhou performed Metropolis Monte Carlo simulations and the calculations of the optical properties. Linwei Zhou coded program "Renmin Mobility Calculator" (ReMoC) for carrier mobility calculations. Ning-Hua Tong carried out the renormalized spin wave theory (RSWT) study of two-dimensional anisotropic Heisenberg model (2DAHM) on square lattice and on honeycomb lattice. Cong Wang, Xieyu Zhou, Ning-Hua Tong, Zhong-Yi Lu and Wei Ji analyzed the results and wrote the manuscript. All authors commented on the manuscript.

Graphical abstract

A category of semiconducting and metallic FM monolayers, namely MnNX and CrCX (X=Cl, Br and I; C=S, Se and Te) were theoretically predicted. Their Curie temperatures are in a range from 108 to 492 K. Eight of these 12 monolayers are semiconducting with their bandgaps varying from nearly 2 to 0 eV. Exceptional large anisotropies of carrier effective mass, mobility and light absorption lead to a novel spin-locked linear dichroism, and carrier motion direction locking.

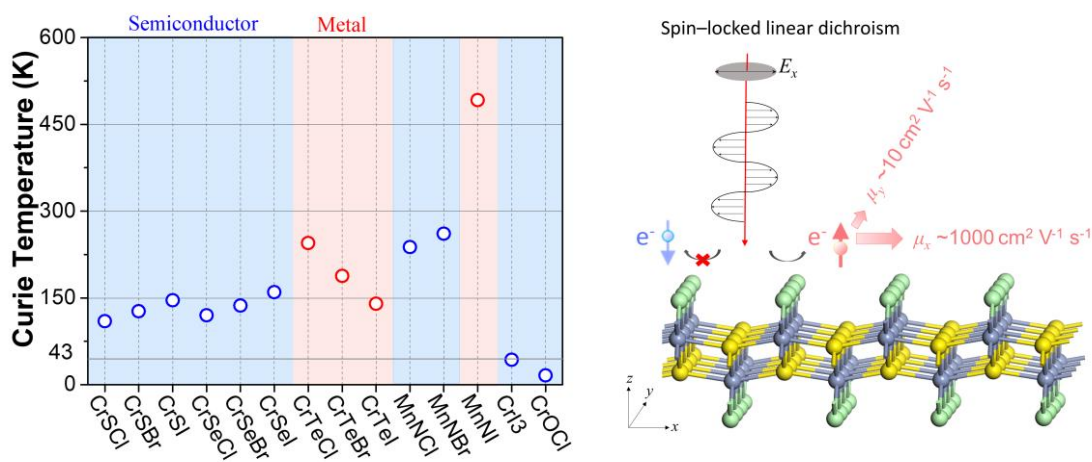

# Supporting Information I

# A family of high-temperature ferromagnetic monolayers with locked spin-dichroism-mobility anisotropy: MnNX and CrCX (X=Cl, Br, I; C=S, Se, Te)

Cong Wang[†], Xieyu Zhou[†], Linwei Zhou, Ning-Hua Tong, Zhong-Yi Lu and Wei Ji[*]

*Beijing Key Laboratory of Optoelectronic Functional Materials & Micro-Nano Devices,
Department of Physics, Renmin University of China, Beijing 100872, P.R. China*

*Email: wji@ruc.edu.cn*

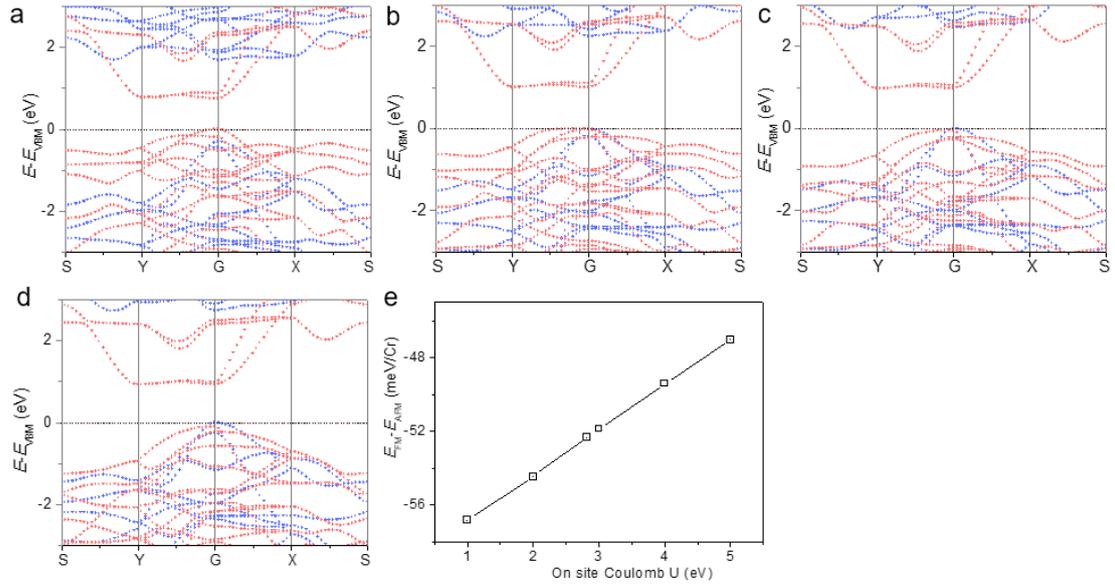

**Supplementary Figure S1. Role of different on-site Coulomb U values.** Electronic band structures of a FM CrSCl monolayer calculated with the *U* values of 1 eV (a), 2 eV (b), 2.8 eV (c) and 4 eV (d), respectively. The effective mass anisotropy and spin up CB1 and CB2 and the positions of CBM and VBM do not appreciably change with respect to increased U values, while the position of the spin-down CBM lifts and the splitting between the spin-up CB1 and CB2 states at G gradually decreases. The U and J values used in our calculations were listed in Table S4, which were derived using a linear response method. Energy differences between the intra-plane FM and sAFM magnetic orders of the CrSCl monolayer as a function of effective U were plotted in (e). It indicates that the FM groundstate is rather robust and is at least 45meV/Cr more stable than other configurations.

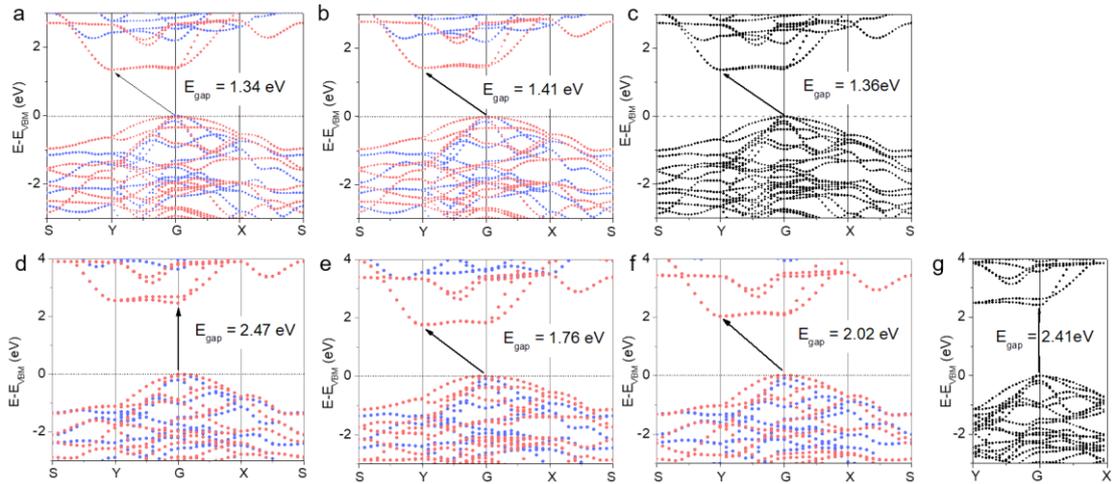

**Supplementary Figure S2. Role of functionals, on-site U, J and SOC in appearance of bandstructures.** Here, we use the CrSBr monolayer as an example to examine the role of DFT functional, U, J and SOC in appearance of bandstructure. Two adding U forms, namely separate U and J values (U-J) and effective U ($U_{eff}$), were discussed. The bandstructures calculated with optB86b-vdW+U-J (a), PBE+U-J (b), PBE+U-J+SOC (c), HSE06+U-J (d), HSE06 (e), HSE06+$U_{eff}$ (f), HSE06+U-J+SOC (g) were plotted. All electronic structure calculations were performed based on the atomic structures fully relaxed using optB86b-vdW. The bandstructures revealed using optB86b-vdW (a) and PBE (b) are essentially the same. By including a portion of exact Hartree-Fock exchange, the HSE06 hybrid functional often gives more reasonable bandgaps, as we primarily discussed in this work. The HSE06 bandgap of CrSBr is 2.47 eV, roughly 1.1 eV larger than that of PBE or optB86b-vdW. The anisotropic feature is rather robust no matter which functional was used. The inclusion of U significantly increases the bandgap from 1.76 eV (e) to 2.02 eV (f) and eventually to 2.47 eV (d). The HSE06+U-J result suggests the CrSBr monolayer a direct bandgap semiconductor while the PBE+U-J, HSE06 and HSE06+$U_{eff}$ calculation do not, which is ascribed to the enlarged splitting between the CB1 and CB2 bands at the G point. The inclusion of exact exchange or consideration of U-J increases the splitting by 139.1, 189.4 and 151.9 meV for the results of PBE+U-J, HSE06 and HSE06+$U_{eff}$ calculations, respectively. This splitting is further enhanced by SOC, i.e. 1.5 meV for PBE+U-J and 3.3 meV for HSE06+U-J. Bandstructures of the all CCX and MnNX monolayers are available in Supplementary Figs. S8 (HSE+U-J), S9 (HSE+U-J+SOC), S10 (PBE+U-J) and S11 (PBE+U-J+SOC), respectively.

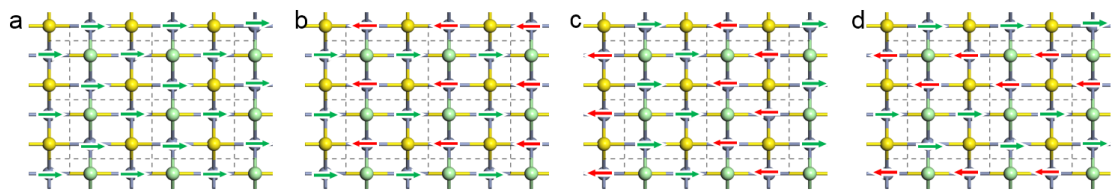

**Supplementary Figure S3. Schematic representation of four magnetic orders used for the derivation of spin-exchange coupling parameters of CrCX monolayers.** Green and red arrows indicate the up and down directions of magnetic moments on Cr atoms, respectively. Spin-exchange coupling parameters were extracted by calculating the total energy differences of these magnetic configurations based on a Heisenberg Model described in the methods section.

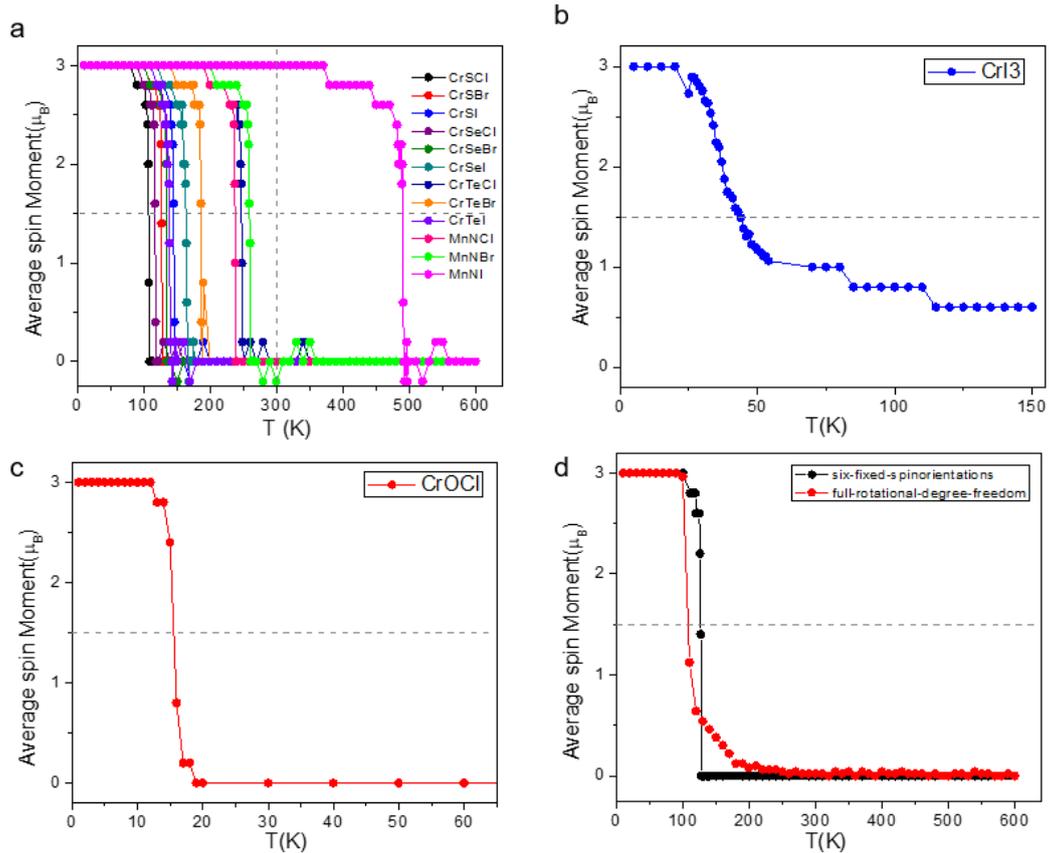

**Supplementary Figure S4. Variation of on-site magnetic moments as a functional of temperature.** Local magnetic moments of Cr or Mn atoms along the easy magnetization axis, as revealed by Monte Carlo simulations, were plotted in (a) for CrCXs and MnNXs, (b) for CrI3 and (c) for CrOCl. (d) the comparison of the simulation of CrSBr using the six-fixed spin orientations model and using the full rotational degree of freedom model. A 2D anisotropic Heisenberg model containing both the easy axis single ion anisotropy and the anisotropic symmetric exchange was adopted with a $50 \times 50$ lattice and the periodic boundary condition. A temperature was recorded as the Curie temperature when the average spin moment becomes lower than half of the initial moment, i.e. 1.5 $\mu_B$ as the gray horizontal dash lines show. A temperature step of 10 K was employed and a much smaller step of 1 K was used around the critical point. The gray vertical dash line indicates the room temperature of 300 K.

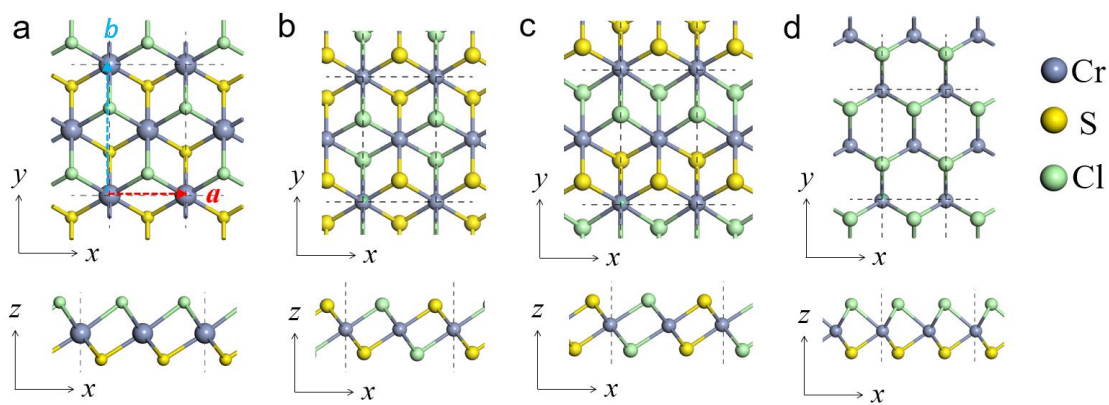

**Supplementary Figure S5. Schematic models of CrCXs with different structure forms.** Top and side views of monolayer CrSCl in hexagonal structures with halogen and chalcogen atoms occupying different sites, namely 1T-janus (a), 1T-aa (b), 1T-ab (c) and 2H-janus (d). Slate-blue, yellow and cyan balls represent Cr, S and Cl atoms, respectively. In-plane ferromagnetic state was found the ground state in each of these layers while the CrCX form shown in Fig. 1 always shows the lowest energy in comparison with these four forms. In addition, we infer that a CrSCl Janus monolayer may tend to bend because of its inherent in-plane strain induced by the different Cr-S and Cr-Cl bond lengths. Detailed structural information and relative total energies of all the configurations were listed in Table S1.

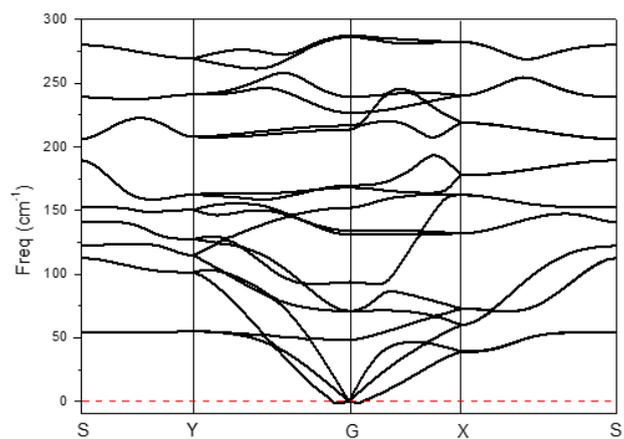

**Supplementary Figure S6.** Phonon dispersion of a CrSeBr monolayer where no appreciable imaginary frequency was found throughout the k-space, suggesting the thermal stability of CrSeBr.

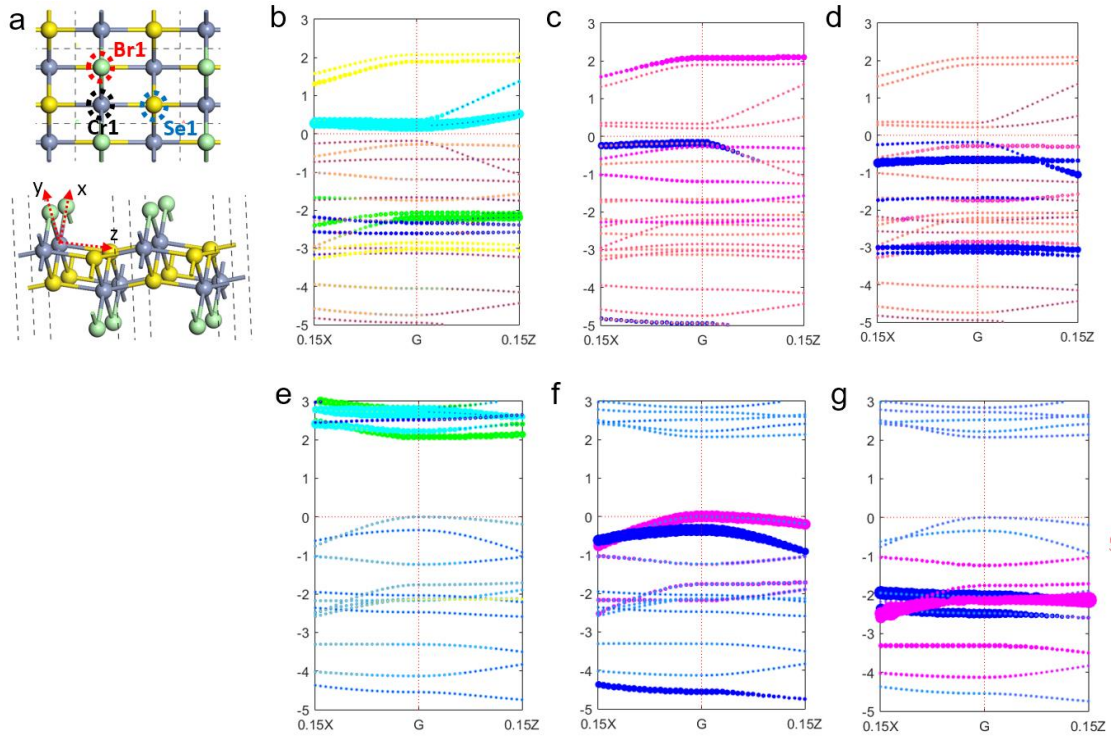

**Supplementary Figure S7. Orbital-decomposed band structures of a ferromagnetic CrSeBr monolayer.** We redefined the axis directions to better illustrate the orbital hybridization, as shown in (a). Electronic band structures of the FM CrSeBr monolayer were highlighted for spin-up (b-d) and spin-down (e-g) components of the Cr, Se and Br atoms marked in (a), respectively. Here, the bandstructure was revealed with the optB86b-vdW functional while other functionals show essentially the same results. The VBM energy is set to zero. Different colors were used to map the Cr-$d$, Se-$p$ and Br-$p$ orbitals, namely green, blue, cyan and yellow for Cr $d_{xy}$, $d_{yz}$/$d_{xz}$ and $d_{z^2}$ and $d_{x^2-y^2}$, respectively; while magenta and blue are for Se or Br $p_x$/$p_y$ and $p_z$, respectively. The hybridization between the Cr-$e_g$ and Se/Br-$p$ orbitals and occupancies are clearly shown in these plots. Note that CrCXs and MnNXs share the similar features in terms of orbital decomposition.

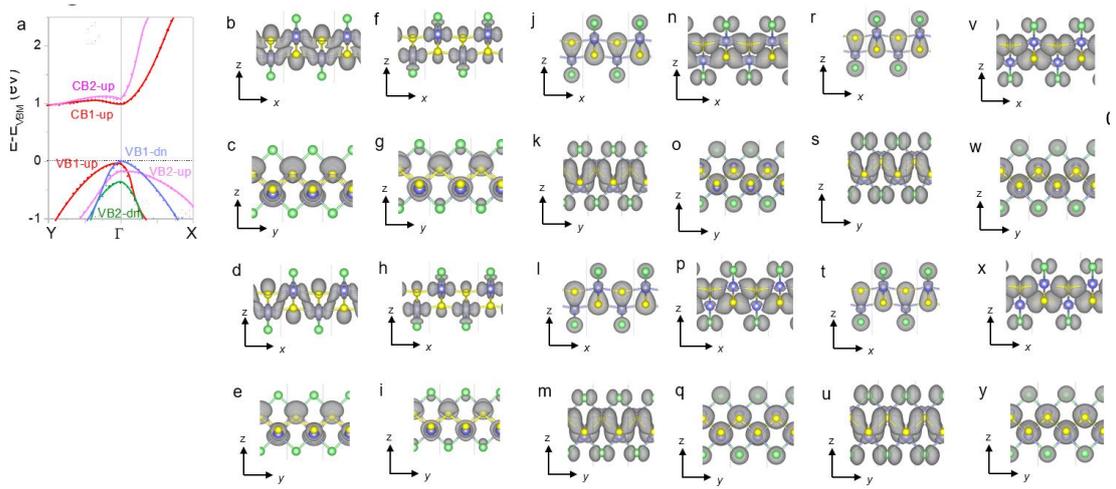

**Supplementary Figure S8. Visualized wavefunctions of the CrSeBr monolayer.** A simplified bandstructure of CrSeBr around the G point was shown in (a). All six relevant bands were named and colored. Side views, plotted in the *x-z* (b, d, f, h, j, l, n, p, r, t, v and x) plane and in the *y-z* plane (c, e, g, I, k, m, o, q, s, u, w and y), of the wavefunction norms of spin-up CB2 along G-X (b, c) and G-Y (d, e), spin-up CB1 along G-X (f, g) and G-Y (h, i), spin-down VB1 along G-X (j, k) and G-Y (l, m), spin-up VB1 along G-X (n, o) and G-Y (p, q), spin-up VB2 along G-X (r, s) and G-Y (t, u) and spin-down VB2 along G-X (v, w) and G-Y (x, y). The isosurface value was set to 0.001 *e*/Bohr3.

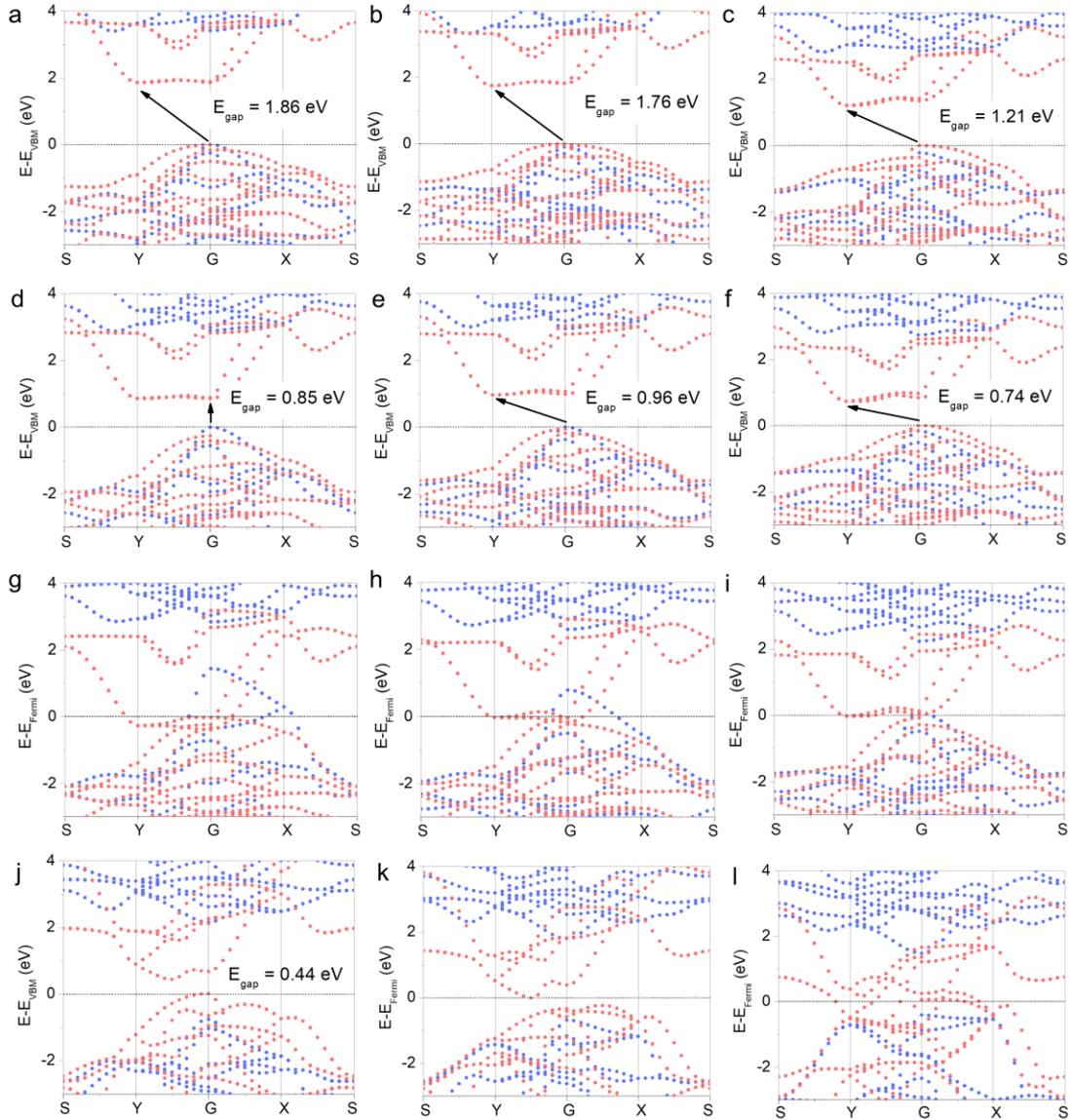

**Supplementary Figure S9. Band structures of the CrCX monolayers.** Electronic band structures of CrSCl (a), CrSBr (b), CrSI (c), CrSeCl (d), CrSeBr (e), CrSeI (f), CrTeCl (g), CrTeBr (h), CrTeI (i), MnNCl (j), MnNBr (k) and MnNI (l) monolayers were calculated using the hybrid HSE06 functional. Among them, CrSCl, CrSBr, CrSeCl and CrSeBr are semiconductors with direct bandgap, which maintains when including spin-orbit coupling. The position of CBM and VBM, relative to the vacuum level, were summarized in Fig. 3. A semiconductor (MnNCl) to semimetal (MnNBr) and finally to metal (MnNI) transition occurs when substituting Cl with Br and I in MnNXs. MnNCl was found a semiconductor with an indirect bandgap of 0.29eV with strong anisotropy along x and y.

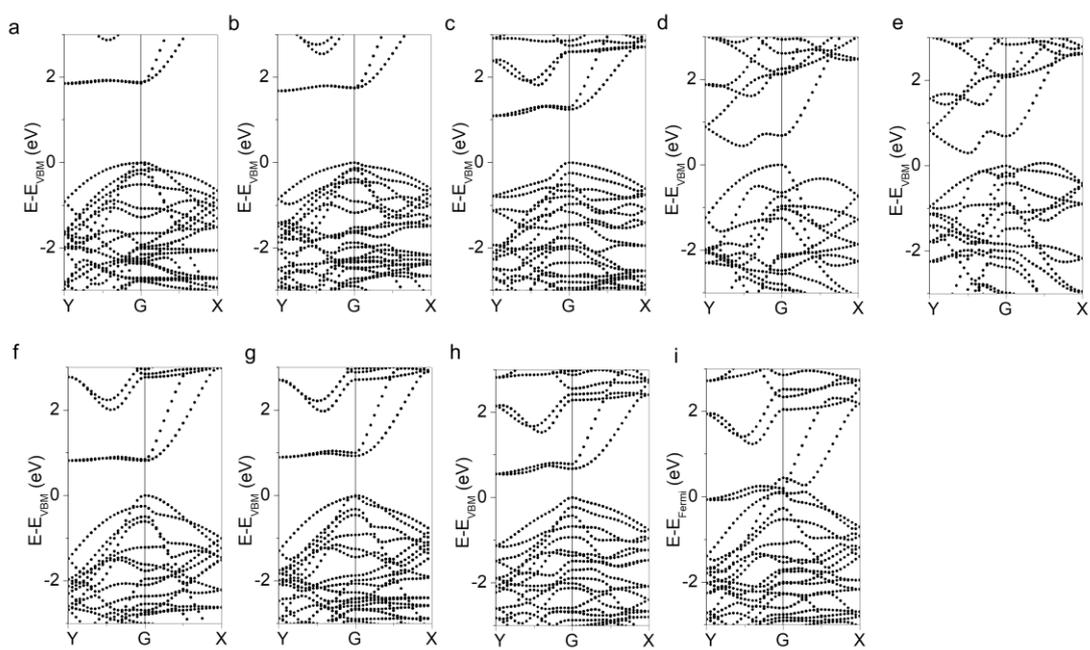

**Supplementary Figure S10. Role of SOC in band structures of CrCX and MnNX monolayers.** The bandstructures of CrSCl (a), CrSBr (b), CrSI (c), MnNCl (d), MnNBr (e), CrSeCl (f), CrSeBr (g), CrSeI (h) and CrTeI (i) were calculated using the HSE06 functional with the inclusion of spin-orbit coupling.

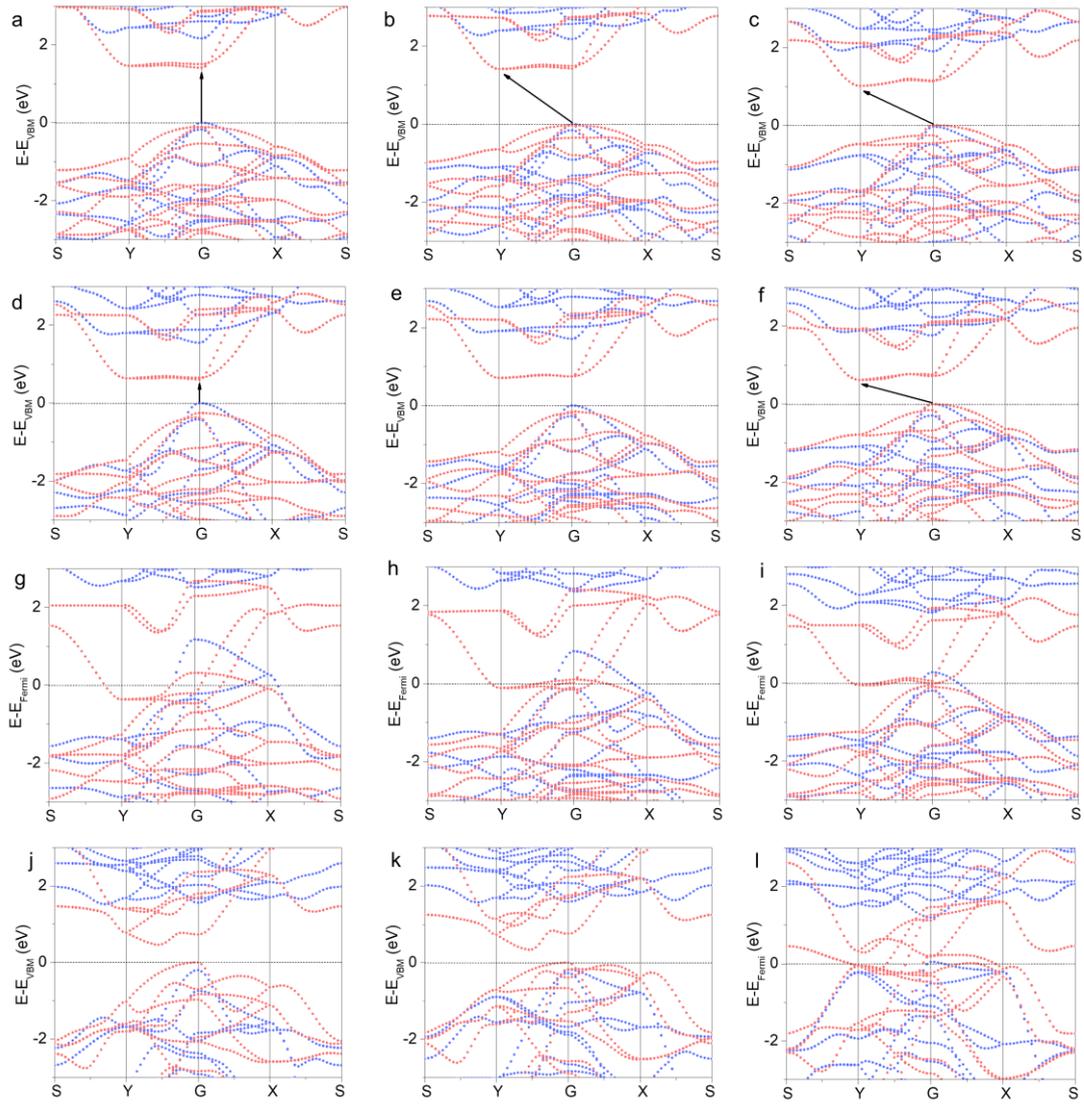

**Supplementary Figure S11. Band structures of monolayer CrCXs calculated with PBE.** Electronic band structures of CrSCl (a), CrSBr (b), CrSI (c), CrSeCl (d), CrSeBr (e), CrSeI (f), CrTeCl (g), CrTeBr (h), CrTeI (i), MnNCl (j), MnNBr (k) and MnNI (l) monolayers were calculated using the PBE functional based on the geometric structures fully optimized using the optB86b-vdW functional. Here, the bandgaps are generally 1 eV smaller than those of the HSE06 results. The Te series were all found metallic while CrTeI was found semiconducting in the HSE06 result.

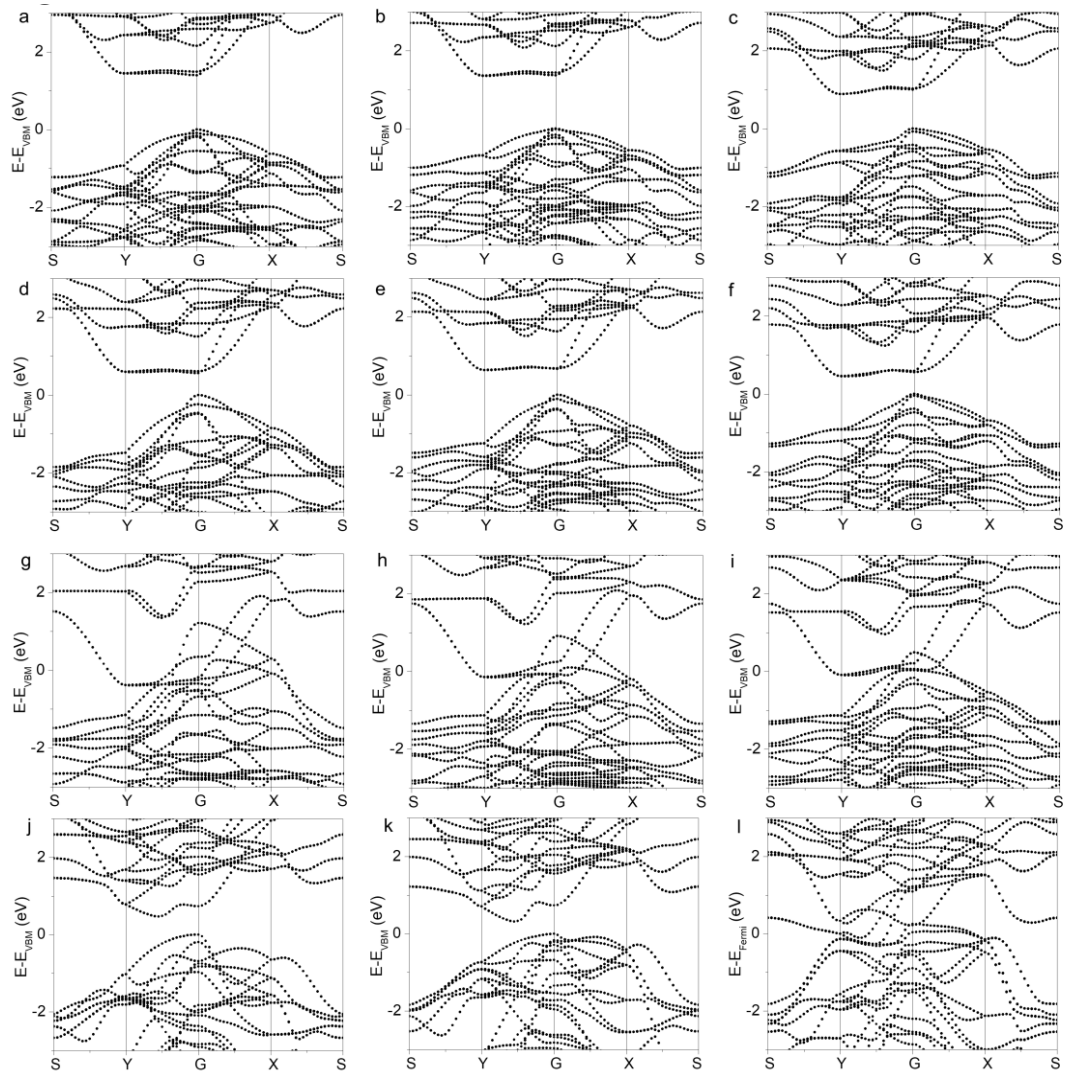

**Supplementary Figure S12. Band structures of CrCX and MnNX monolayers calculated with PBE-SOC.** Electronic band structures of CrSCl (a), CrSBr (b), CrSI (c), CrSeCl (d), CrSeBr (e), CrSeI (f), CrTeCl (g), CrTeBr (h), CrTeI (i), MnNCl (j), MnNBr (k) and MnNI (l) monolayers were calculated using the PBE functional with inclusion of SOC based on the geometric structures fully optimized using the optB86b-vdW functional .

**Table S1**

On-site $U$ and exchange $J$ values of CrCX, MnNX and CrI$_3$ monolayers. These values were calculated using a linear response method, as implemented in the QUANTUM ESPRESSO package.

| Monolayers | $U$ (eV) | $J$ (eV) |
|:---:|:---:|:---:|
| CrSCl  | 4.05 | 0.96 |
| CrSBr  | 4.03 | 0.96 |
| CrSI   | 4.01 | 0.97 |
| CrSeCl | 4.11 | 0.96 |
| CrSeBr | 4.11 | 0.96 |
| CrSeI  | 4.08 | 0.96 |
| CrTeCl | 4.40 | 0.80 |
| CrTeBr | 4.38 | 0.80 |
| CrTeI  | 4.25 | 0.89 |
| CrOCl  | 3.89 | 0.98 |
| MnNCl  | 4.29 | 1.23 |
| MnNBr  | 4.24 | 1.24 |
| MnNI   | 4.13 | 1.19 |
| CrI$_3$ | 3.90 | 1.10 |

**Table S2**

Relative total energy $\Delta E_0$ to the most stable configuration calculated with the same on site Coulomb U and J), magnetic moment ($M$), lattice constants $a$ and $b$ of the CrSCl monolayer. Here, these results of CrSCl were used as illustrations while all ferromagnetic orders in the CrCX structure are energetically favored in all CrCX and MnNX monolayers, which is consistent with our previous results of the 1$e$/Cr doped CrS$_2$ monolayer. The CrCX structure form is at least 0.1eV/Cr more stable than these configurations.

| CrSCl-1L | $\Delta E_0$ (eV/Cr) | $M$ (μ$_B$/Cr) | $a$ (Å) | $b$ (Å) |
| --- | --- | --- | --- | --- |
| CrCX-FM | 0.000 | 2.93 | 4.80 | 3.44 |
| CrCX-sAFM | 0.029 | 2.86 | 4.81 | 3.44 |
| 1T-janus-FM | 0.131 | 3.00 | 3.44 | 5.96 |
| 1T-janus-ab | 0.170 | 2.84 | 3.44 | 5.91 |
| 1T-aa-FM | 0.430 | 2.94 | 3.48 | 5.99 |
| 1T-aa-sAFM | 0.445 | 2.83 | 3.47 | 5.93 |
| 1T-ab-FM | 0.127 | 2.99 | 3.45 | 6.09 |
| 1T-ab-sAFM | 0.145 | 2.86 | 3.45 | 6.02 |
| 2H-janus-FM | 0.971 | 3.08 | 3.35 | 5.81 |
| 2H-janus-sAFM | 0.981 | 2.88 | 3.34 | 5.74 |

**Table S3**

Bond lengths (*d*) and angles of FM CrCX monolayers. In-plane magnetic order does not significantly affect the geometric structure of the monolayer, different from the case found in the CrS$_2$ monolayer. Substitution of S atoms with Se or Te atoms significantly enlarges the bond lengths and lattice constants along the *x* and *y* directions, since S atoms bridge Cr atoms in both Cr-S/Cl-Cr rhomboid chains along the *y* direction and distorted Cr-S-Cr rectangular chains along the *x* direction. Replacement of halogen atoms only tunes the structure in the y direction. The Cr-C-Cr–$J_1$ angles along the rhomboid chains increase when substituting the S atom with Se or Te while the Cr-X-Cr-$J_1$ angles show an opposite trend.

| Monolayers | d-$J_1$ (Å) | d-$J_2$ (Å) | d-$J_3$ (Å) | d-Cr-C-$J_3$ (Å) | ∠Cr-X-Cr-$J_1$ (°)[1] | ∠Cr-C-Cr-$J_1$ (°)[1] | ∠Cr-C-Cr-$J_2$ (°)[2] |
|---|---|---|---|---|---|---|---|
| CrSCl | 3.44 | 3.61 | 4.80 | 2.43 | 92.62 | 91.49 | 96.59 |
| CrSBr | 3.54 | 3.61 | 4.78 | 2.43 | 88.91 | 94.22 | 96.61 |
| CrSI | 3.69 | 3.63 | 4.76 | 2.42 | 85.47 | 98.47 | 96.61 |
| CrSeCl | 3.52 | 3.79 | 5.09 | 2.57 | 94.81 | 88.50 | 96.22 |
| CrSeBr | 3.62 | 3.80 | 5.07 | 2.57 | 91.07 | 91.08 | 96.40 |
| CrSeI | 3.78 | 3.82 | 5.05 | 2.56 | 87.55 | 95.29 | 96.53 |
| CrTeCl | 3.55 | 4.17 | 5.65 | 2.85 | 95.92 | 79.77 | 95.62 |
| CrTeBr | 3.71 | 4.15 | 5.57 | 2.82 | 93.36 | 84.52 | 96.31 |
| CrTeI | 3.91 | 4.14 | 5.50 | 2.79 | 90.53 | 90.30 | 96.65 |
| MnNCl | 3.09 | 2.94 | 3.84 | 1.96 | 83.02 | 103.04 | 96.94 |
| MnNBr | 3.20 | 2.95 | 3.84 | 1.96 | 80.16 | 106.41 | 96.41 |
| MnNI | 3.41 | 2.95 | 3.86 | 1.96 | 77.14 | 113.53 | 95.10 |

[1] Imaging Cr atom along the *y* ($J_1$) direction.

[2] Another Cr atom along the $J_2$ direction.

**Table S4**

Isotropic spin-exchange coupling parameters ($J_1$-$J_3$), single ion easy axis anisotropy ($\lambda$) and anisotropic symmetric spin-spin coupling parameter ($D$) of CrTeI derived using the PBE+UJ, PBE, HSE+UJ and HSE methods. We use CrTeI as an example to show the influence of functional and U values on the derived spin-exchange coupling parameters and thus the predicted Curie temperatures. Our results show that both the spin-exchange coupling and magnetic anisotropies of the CrTeI monolayer are significantly enhanced in HSE calculations, thus leading to an even higher $T_c$ of 400 K as predicted using the AH model.

| CrTeI-1L | $J_1$ (meV) | $J_2$ (meV) | $J_3$ (meV) | $\lambda$ (meV) | $D$ (meV) | axis | $T_c$ (K) |
|---|---|---|---|---|---|---|---|
| PBE+UJ | 4.56 | 4.13 | -1.90 | -0.104 | 0.552 | z | 139 |
| PBE | 5.17 | 8.17 | 4.37 | -0.118 | 0.578 | z | 361 |
| HSE+UJ | 5.51 | 6.88 | -1.86 | 0.333 | 0.439 | x | |
| HSE | 6.42 | 9.63 | -0.13 | 0.331 | 2.038 | z | 401 |

**Table S5**

Phonon-limited electron/hole mobility of the FM CrSeBr monolayer, compared with those of MoS$_2$, PtSe$_2$ and black phosphorus. Here, $m_x^*$ and $m_y^*$ are carrier effective masses along the $x$ and $y$ directions, respectively. Variables $E_{1x}$ ($E_{1y}$), $C_{x\_2D}$ ($C_{y\_2D}$), and $\mu_{x\text{-}2D}$ ($\mu_{y\text{-}2D}$) are the deformation potentials, 2D elastic modulus, and the mobility along the $x$ ($y$) direction, respectively.

| Carrier type | | $m_x^*/m_0$ | $m_y^*/m_0$ | $C_{x\_2D}$ | $C_{y\_2D}$ | $E_{1x}$ | $E_{1y}$ | $\mu_{x\text{-}2D}$ | $\mu_{y\text{-}2D}$ |
|---|---|---|---|---|---|---|---|---|---|
| | | | | (J m$^{-2}$) | | (eV) | | ($10^3$ cm$^2$ V$^{-1}$ s$^{-1}$) | |
| e | dn-CB1 | 1.51 | 0.34 | 91.18 | 73.78 | 1.59 | 3.66 | 0.32 | 1.02 |
| | up-CB2 | 0.06 | 1.18 | 91.18 | 73.78 | 9.22 | 3.43 | 6.10 | 0.09 |
| | up-CB1 | 0.40 | 7.14 | 91.18 | 73.78 | 4.45 | 0.18 | 0.61 | 0.79 |
| | MoS$_2$ | 0.47 | 0.47 | 129.93 | 135.34 | 5.57 | 5.55 | 0.41 | 0.43 |
| | PtSe$_2$ | 0.26 | 0.48 | 66.95 | 66.82 | 2.20 | 0.72 | 3.25 | 16.25 |
| | BP | 0.17 | 1.12 | 14.5 | 50.8 | 2.72 | 7.11 | 0.56 | 0.04 |
| h | dn-VB1 | 0.57 | 0.24 | 91.18 | 73.78 | 1.40 | 5.65 | 3.05 | 0.85 |
| | up-VB1 | 0.06 | 1.30 | 91.18 | 73.78 | 1.69 | 4.97 | 93.89 | 0.14 |
| | up-VB2 | 2.66 | 0.53 | 91.18 | 73.78 | 1.00 | 2.55 | 0.14 | 0.23 |
| | dn-VB2 | 0.18 | 0.86 | 91.18 | 73.78 | 1.40 | 1.01 | 3.05 | 2.07 |
| | BP | 0.15 | 6.35 | 14.47 | 50.80 | 2.50 | 0.15 | 0.33 | 7.71 |